\documentclass[aps,pra,floatfix,showpacs,twocolumn,tightenlines,superscriptaddress]{revtex4}

\usepackage{amsmath}
\usepackage{graphicx}
\usepackage{amsfonts}

\begin{document}

\title{Quantum Control of a Single Qubit}

\author{Agata M.~Bra{\'n}czyk}
\author{Paulo E. M. F.~Mendon{\c c}a}
\author{Alexei Gilchrist}
\author{Andrew C. Doherty}
\affiliation{Department of Physics, The University of Queensland,
Queensland 4072, Australia}
\author{Stephen D. Bartlett}
\affiliation{School of Physics, The University of Sydney, New South
Wales 2006, Australia}
\date{14 August 2006}

\begin{abstract}
Measurements in quantum mechanics cannot perfectly distinguish all
states and necessarily disturb the measured system. We present and
analyse a proposal to demonstrate fundamental limits on quantum
control of a single qubit arising from these properties of quantum
measurements.  We consider a qubit prepared in one of two
non-orthogonal states and subsequently subjected to dephasing noise.
The task is to use measurement and feedback control to attempt to
correct the state of the qubit.  We demonstrate that projective
measurements are not optimal for this task, and that there exists a
non-projective measurement with an optimum measurement strength
which achieves the best trade-off between gaining information about
the system and disturbing it through measurement back-action. We
study the performance of a quantum control scheme that makes use of
this weak measurement followed by feedback control, and demonstrate
that it realises the optimal recovery from noise for this system. We
contrast this approach with various classically inspired control
schemes.
\end{abstract}


\maketitle
\section{Introduction}
\label{sec:intro}

Any practical quantum technology, such as quantum key distribution
or quantum computing, must function robustly in the presence of
noise.  Many modern ``classical'' technologies tolerate noise,
faulty parts, etc., by relying on \emph{feedback control} systems,
which monitor the system and use this information to control its
state.  Given the ubiquity and power of feedback control for
classical systems, it is worthwhile investigating how such control
concepts can be applied to quantum technologies as well.  However,
strategies for \emph{quantum control} must take into account
some fundamental features of quantum mechanics, namely, restrictions
on information gain, and measurement back-action.

Classically, it is possible in principle to acquire all the
information about the state of a system with certainty by using
sufficiently precise measurements.  That is, the state of a single
classical system can be precisely determined via measurement. For
quantum systems, however, this is not always possible: if the system
is prepared in one of several non-orthogonal states, no measurement
can determine which preparation occurred with certainty.

In addition, for quantum systems, monitoring comes at a price: any
measurement that acquires information about a system must
necessarily disturb it uncontrollably.  This feature is often
referred to as \emph{back-action} --- the fundamental noise induced
on a system through any measurement, which  maintains the
uncertainty relations.  This feature of quantum measurement is also
distinct from the classical situation, wherein measurements that do
not alter the state of the system can in principle be performed.

These two fundamental features of quantum systems --- that
non-orthogonal states cannot be perfectly discriminated, and that
any information gain via measurement necessarily implies disturbance
to the system --- require a reevaluation of conventional methods and
techniques from control theory when developing the theory of quantum
control.

In this paper, we investigate the use of measurement and feedback
control of a single qubit, prepared in one of two non-orthogonal
states and subsequently subjected to noise.  Our main result is
that, in order to optimize the performance of the control scheme (as
quantified by the average fidelity of the corrected state compared
to the initial state), one must use non-projective measurements with
a strength that balances the trade-off between information gain and
disturbance.

Belavkin was the first to recognise the importance of feedback
control for quantum systems and describe a theoretical framework for
analysing both discrete and continuous time models
\cite{83belavkin178,belavkin99a}. Despite this early start, it is
only recently that the degree of control and isolation of quantum
systems has progressed to the point that the experimental
exploration of quantum control tasks has been possible
\cite{02armen133602,02smith133601,04geremia270,04reiner023819,04lahaye74,06bushev043003},
and the field is now undergoing rapid development (see for example
\cite{05job}).

The specific control problem we are interested in here is the
stabilization against noise of states of a single two level system.
Similar problems have been considered in continuous time feedback
models, e.g., the stabilization of a single state of a driven and
damped two-level atom~\cite{wang02a,02wiseman013807} and the
maintenance of the coherence of a qubit undergoing
decoherence~\cite{lidar05}. Several recent papers have investigated
state preparation and feedback stabilization onto eigenstates of a
continuously-measured observable in higher-dimensional
systems~\cite{vanhandel05a,mirrahimi05a}.  In contrast to these
prior investigations, we investigate a feedback scheme to stabilize
\emph{two} non-orthogonal states of a two-level system.  We work in
a discrete-time setting, rather than continuous-time as considered
in most prior work, which considerably simplifies the problem and
most clearly illustrates the central concepts.  Gregoratti and
Werner have investigated exactly this kind of model of recovering
the state of the system after interaction with the
environment~\cite{gregoratti03a,gregoratti04a} in the case where it
is possible to make measurements on the environment. In our setting
we imagine that the environment that causes the initial decoherence
is not available subsequently for the feedback protocol. Our main
interest is to investigate the effects of the kind of trade-off
between information and disturbance that is ubiquitous in quantum
information in a concrete optimal control problem. Related
information-disturbance trade-offs in quantum feedback control are
discussed in~\cite{jacobs01}. Finally, we note that implementing
quantum operations on a single qubit through the use of measurement
and feedback control as considered here has been investigated for
eavesdropping strategies in quantum cryptography~\cite{Niu99} and
for engineering general open-system dynamics~\cite{Llo01}.

Note that there is a fundamental difference between the kind of
quantum control problem we are considering here and the related task
of quantum error correction. (For an introduction to the latter,
see~\cite{Nie00}.)  The essence of quantum error correction is to
\emph{encode} abstract quantum information into a physical quantum
system and to choose degrees of freedom that are unaffected by the
relevant noise, or upon which errors can be deterministically
corrected. However, it can be the case that one wishes to protect
particular physical degrees of freedom of quantum systems and one is
not free to choose an arbitrary encoding.  (One such example is
\emph{reference frame distribution} via the exchange of quantum
systems). The quantum states required for these schemes cannot be
encoded into quantum error correcting codes or noiseless
subsystems~\cite{Pre00}; protecting such systems from noise may
therefore be an application of this kind of quantum control.

The paper is structured as follows.  In section~\ref{sec:task}, we
define the control task in detail; in section~\ref{sec:Classical},
we present and determine the performance of control strategies based
on ``classical'' concepts. Section~\ref{sec:quantum} introduces our
quantum strategy, investigating the use of \emph{weak} quantum
measurements, and analyses its performance against the strategies of
section~\ref{sec:Classical}.  We also demonstrate that our quantum
control scheme is optimal for the task at hand.  In
section~\ref{sec:conclusions} we discuss the implications of our
result and their relevance to other problems.

\section{A Simple Control Task}
\label{sec:task}

The aim of this paper is to explore the key issues we will confront
when applying concepts from control theory to finite-dimensional
quantum systems.  In order to facilitate the analysis and to be able
to concentrate on the key departures from classical control, we will
chose a very simple quantum system and noise model.  The emphasis is
not towards a practical task, but as an illustrative example.

\begin{figure}
  \begin{center}
   \includegraphics[width=0.46\textwidth]{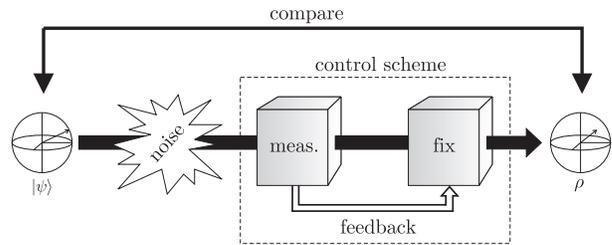}
  \end{center}
  \caption[Schematic of a quantum control procedure.]
  {Schematic of a quantum control procedure. A qubit,
  subjected to dephasing noise, is subsequently measured and corrected
  based on the results of this measurement. The output state
  $\rho$ is compared with the input state $|\psi\rangle$ to
  characterise how well the scheme performs.}
  \label{fig:compare}
\end{figure}

Consider the following operational task: a qubit prepared in one of two
non-orthogonal states $|\psi_1\rangle$ or $|\psi_2\rangle$ (with overlap
$\langle\psi_1|\psi_2\rangle = \cos\theta$ for $0\leq \theta \leq \pi/2$) is
transmitted along a noisy quantum channel. Without knowing which state was
transmitted, we will attempt to ``correct'' the system, i.e., undo the effect
of the noise, through the use of a control scheme based on measurement and
feedback; see Fig.~\ref{fig:compare}.

The noise model that we will consider is \emph{dephasing noise}.
Let $\{|0\rangle,|1\rangle\}$ be a basis for
the qubit Hilbert space, and the Pauli operator $Z$ is the unitary
operator defined by $Z|0\rangle = |0\rangle$, $Z|1\rangle = -
|1\rangle$.  Dephasing noise is characterized as follows: with
probability $p$ a phase-flip $Z$ is applied to the system, and with
probability $1-p$ the system is unaltered.  The noise is thus
described by a quantum operation~\cite{Nie00}, i.e., a
completely-positive trace-preserving (CPTP) map $\mathcal{E}_p$,
that acts on a single-qubit density matrix $\rho$ as
\begin{equation}\label{eq:erho}
  \mathcal{E}_p(\rho)=p(Z\rho Z)+(1-p)\rho\,.
\end{equation}
We will consider the noisy channel to be fully characterized, meaning that $p$
is known and without loss of generality in the range $0\leq p \leq 0.5$.

We will choose the two initial states to be oriented in such a way that their
distinguishability, as measured by their trace distance, is maintained under
the action of the noise.  It is straightforward to show that this condition is
satisfied by the states
\begin{align}
  \label{eq:State1}
  |\psi_1\rangle&=\cos\tfrac{\theta}{2}|{+}\rangle+\sin\tfrac{\theta}{2}|{-}\rangle\,,\\
  \label{eq:State2}
  |\psi_2\rangle&=\cos\tfrac{\theta}{2}|{+}\rangle-\sin\tfrac{\theta}{2}|{-}\rangle\,,
\end{align}
where $|{\pm}\rangle=(|0\rangle\pm |1\rangle)/\sqrt{2}$.

\begin{figure}
  \begin{center}
   \includegraphics[width=0.25\textwidth]{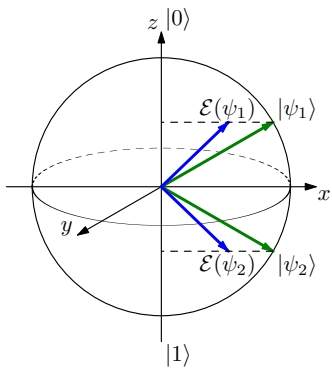}
  \end{center}
  \caption[Initial states on Bloch sphere]
  {Bloch sphere representation of the initial states, and the states after the noise.
  The noise shortens the Bloch vectors along the $x$-axis. We have used the notation
  $\mathcal{E}(\psi)$ as a shorthand for $\mathcal{E}(|\psi\rangle\langle\psi|)$.}
  \label{fig:bloch}
\end{figure}

Consider the Bloch sphere defined by states $|0\rangle$ and
$|1\rangle$ as the poles on the $z$-axis.   The two states
$|\psi_1\rangle$ and $|\psi_2\rangle$ lie in the $x{-}z$ plane and
straddle the equator of the Bloch sphere by angles $\pm\theta$; see
Fig.~\ref{fig:bloch}. On this Bloch sphere, the dephasing noise
acting on these states has the effect of decreasing the
$x$-component of their Bloch vectors. The trace distance between
these two states, given by the Euclidean distance between their
Bloch vectors, is invariant under this dephasing noise.

We now consider whether there exists a control procedure
$\mathcal{C}$ (some ``black box'') that can correct the state of
this system and counteract the noise, at least to some degree,
independent of which input state was prepared.  To quantify the
performance of any such procedure, we will use the average fidelity
to compare the noiseless input states $|\psi_i\rangle$ with the
corrected output states $\rho_i$. Assuming an equal probability for
sending either state $|\psi_1\rangle$ or $|\psi_2\rangle$, the
figure of merit is
\begin{align}\label{eq:AverageFidelity}
    F_{\mathcal{C}} &= \tfrac{1}{2}F(|\psi_1\rangle,\rho_1) +
    \tfrac{1}{2}F(|\psi_2\rangle,\rho_2) \nonumber \\
    &= \tfrac{1}{2}\langle \psi_1 |\rho_1 |\psi_1\rangle +
    \tfrac{1}{2}\langle \psi_2 |\rho_2 |\psi_2\rangle \,,
\end{align}
where the fidelity between a pure state $|\psi\rangle$ and a mixed
state $\rho$ is defined as $F(|\psi\rangle,\rho) \equiv
\langle\psi|\rho|\psi\rangle$.  The fidelity $F$ ranges from 0 to 1
and is a measure of how much two states overlap each other (a
fidelity of 0 means the states are orthogonal, whereas a fidelity of
1 means the states are identical). It has the following simple
operational meaning when the input state is pure: the fidelity
$F(|\psi_i\rangle,\rho_i)$ is the probability that the state
$\rho_i$ will yield outcome $|\psi_i\rangle$ from the projective
measurement
$\{|\psi_i\rangle\langle\psi_i|,|\psi_i^\perp\rangle\langle\psi_i^\perp|
\}$.

Thus, the aim is to find a control operation, described by a CPTP
map $\mathcal{C}$ independent of the choice of initial state, such
that the corrected states
\begin{equation}\label{eq:CorrectedState}
    \rho_i =
    \mathcal{C}\bigl[\mathcal{E}_p(|\psi_i\rangle\langle\psi_i|)\bigr] \,,
\end{equation}
for $i=1,2$ are close to the original states as quantified by the
average fidelity.  We consider control operations that consist of
two steps:  a measurement on the quantum system, followed by a
feedback operation that is conditioned on the measurement result, as
shown in Fig.~\ref{fig:compare}.

\section{Classical Control}
\label{sec:Classical}

In this section, we introduce two types of control schemes for this
task, both of which are based on classical concepts, and we
calculate the performance of these schemes based on the average
fidelity.  In Sec.~\ref{sec:quantum}, we will introduce a quantum
control scheme that outperforms both of these classical schemes.

\subsection*{Classical Strategy A:  Discriminate and Reprepare}

For the control of classical systems, it is always advantageous to
acquire as much information about the system as possible in order to
implement the best feedback scheme.  In line with this principle, a
possible control strategy would be to perform a measurement on the
system which attempts to discriminate between the input states, and
then to reprepare the system in some state based on the measurement
result.

We first characterize all possible discriminate-and-reprepare
schemes; such schemes are associated with \emph{entanglement
breaking trace preserving} (EBTP) maps~\cite{HSR03,Rus03}, as
follows. Any discrimination step is described by a generalized
measurement, (or positive operator-valued measure (POVM))~\cite{Nie00}
yielding a classical probability distribution. The generalized
measurement is
described by the operators $\{P_a\}$ with $P_a\geq 0$ and
$\sum_a P_a = I$.
The resulting map on the quantum system is called a
\emph{quantum-classical} map $QC$~\cite{Hol98}, given by
\begin{equation}
  QC(\rho)=\sum_a{{\rm Tr}\left[\rho P_a\right]|e_a\rangle\langle e_a|}\,,
\end{equation}
where $\{ |e_a\rangle \}$ is an orthonormal basis.  The reprepare
step, in which the quantum system is re-prepared based on the
classical measurement outcome, is described by a
\emph{classical-quantum} map $CQ$~\cite{Hol98}, given by
\begin{equation}
  CQ(\rho)=\sum_{b}{{\rm Tr}\left[\rho |e_b\rangle\langle e_b|\right]Q_b}\,,
\end{equation}
where $\{Q_b\}$ are density matrices.

The concatenation $(CQ \circ QC)(\rho)$ leads to a map of the
form
\begin{equation}\label{eq:EBTP}
  \mathcal{B}(\rho) =\sum_b{{\rm Tr}\left[\rho P_b\right]Q_b}\,.
\end{equation}
This map is an entanglement breaking channel. The name arises
because the output system is unentangled with any other system,
regardless of its input state.  In fact it is straightforward to see
from~\cite{HSR03,Rus03} that all EBTP maps can be realised by some
discriminate-and-reprepare scheme. Thus these EBTP maps formalize
our notion of discriminate-and-reprepare strategies.

The measurement for discriminating two (possibly mixed) preparations
given by Helstrom~\cite{Hel76} is optimal in terms of maximizing the
average probability of a success.  For our choice of states,
Helstrom's measurement is a projective measurement onto the basis
$\{ |0\rangle, |1\rangle \}$, which successfully discriminates the
states $|\psi_1\rangle$ and $|\psi_2\rangle$ with probability
$P_{\rm Helstrom} = \frac{1}{2}(1+\sin\theta)$. Note that because of
the particular choice of dephasing noise, this success probability is
independent of the noise strength $p$.

We now present and analyse two possible discriminate-and-reprepare
strategies, both of which are based on Helstrom's measurement.

\subsubsection*{Discriminate and Reprepare Scheme 1:}

With the outcome of Helstrom's measurement, one strategy is to
reprepare the qubit in either state $|\psi_1\rangle$ or
$|\psi_2\rangle$ based on this measurement outcome.  This scheme
yields an average fidelity of
\begin{equation}
  F_{\rm DR1} =
  1-\tfrac{1}{2}\left(\sin^2{\theta}-\sin^3{\theta}\right)\,.
  \label{eq:fcl}
\end{equation}
Such a replacement ignores the fact
that the discrimination step can fail, with probability $1-P_{\rm
Helstrom}$, in which case a prepared state $|\psi_1\rangle$ would be
reprepared as $|\psi_2\rangle$ (or vice versa).

\subsubsection*{Discriminate and Reprepare Scheme 2:}

\begin{figure*}
\begin{center}
\includegraphics[width=\textwidth]{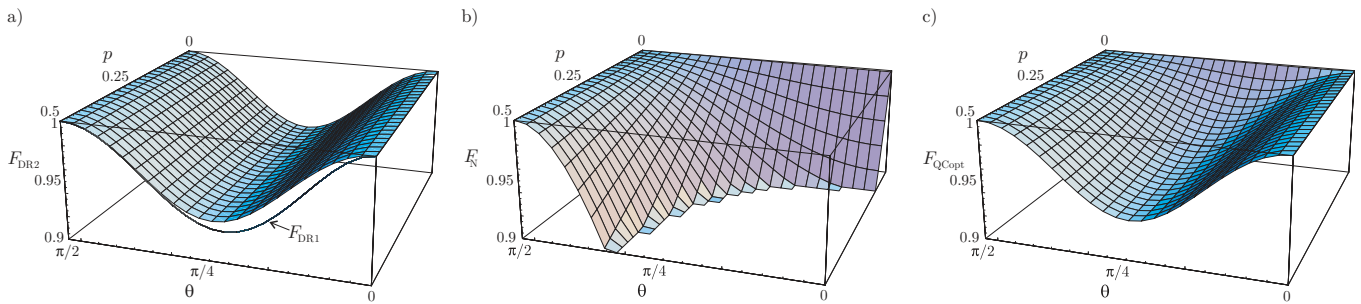}
\end{center}
\caption{The performance of the schemes, quantified by the average
fidelity, as a function of the amount of noise $p$ and the angle
between the input states $\theta$. a) Discriminate and reprepare
scheme quantified by the average fidelity $F_{\rm DR2}$ of
Eq.~(\ref{eq:fcl2}). The fidelity $F_{\rm DR1}$ of
Eq.~(\ref{eq:fcl}) is shown as a solid line at $p=0.5$. Both average
fidelities $F_{\rm DR2}$ and $F_{\rm DR1}$ are independent of $p$.
b)``Do nothing'' scheme, quantified by the average fidelity $F_{\rm
N}$ of Eq.~(\ref{eq:fn}). For this scheme, the average fidelity
drops to $F_\textrm{N} = 1/2$ for $p=0.5$ and $\theta=0$. c) Quantum
control scheme, quantified by the average fidelity
$F_{\mathrm{QCopt}}$ of Eq.~(\ref{eq:fid_qc}). The range of
fidelities plotted has been made identical in all the figures to aid
comparison. }
 \label{fig:fid_all}
\end{figure*}

We can consider other strategies that reprepare different states so
as to reduce the effect of the aforementioned error.  In particular,
we now demonstrate that the following pair of states maximises the
average fidelity:
\begin{equation}
  |\Psi_\pm\rangle=\sqrt{\tfrac{1}{2}\pm\tfrac{\sin^2{\theta}}{2\gamma}}
  |0\rangle+\sqrt{\tfrac{1}{2}\mp\tfrac{\sin^2{\theta}}{2\gamma}}|1\rangle\,,
\end{equation}
where $\gamma\equiv\sqrt{\sin^4{\theta}+\cos^2{\theta}}$. Note that
this replacement is also independent of $p$. Here, $|\Psi_+\rangle$
is prepared if the measurement outcome corresponds to
$|\psi_1\rangle$, and $|\Psi_-\rangle$ is prepared otherwise. In
this strategy, the reprepared states are slightly biased towards the
alternate state to that suggested by the measurement (smaller
$\theta$) --- in a sense hedging our bet. As a proof of the
superiority of this scheme over the former, the fidelity
\begin{equation}
  F_{\rm DR2} =
  \frac{1}{2}+\frac{1}{2}\sqrt{\cos^2{\theta}+\sin^4{\theta}}\,,
  \label{eq:fcl2}
\end{equation}
satisfies $F_{\rm DR2}\geq F_{\rm DR1}$ for all $\theta$. Both
$F_{\rm DR1}$ and $F_{\rm DR2}$ are presented in
Fig.~\ref{fig:fid_all}(a).

This second discriminate-and-reprepare scheme is in fact the
\emph{optimal} discriminate-and-reprepare scheme, in that it
achieves the highest average fidelity
\begin{equation}
  \max_{\mathcal{B}} F_\mathcal{B}
  = \max_{\mathcal{B}} \tfrac{1}{2}\sum_{i=1}^{2}{\langle\psi_i|
    \mathcal{B}\bigl[\mathcal{E}_p(|\psi_i\rangle\langle\psi_i|)\bigr]|\psi_i\rangle}
    \,,
  \label{eq:EBTPoptimisation}
\end{equation}
where the maximization is over all EBTP maps $\mathcal{B}$ acting on
a single qubit.  This optimization was performed (in a different
setting) by Fuchs and Sasaki~\cite{Fuc03}. In the Appendix, we
provide an alternate proof of optimality using techniques from convex
optimization.

\subsection*{Classical Strategy B:  Do Nothing}

Another control strategy would be to do nothing to correct the
states.  Although trivial, this strategy is of interest for
comparison with other schemes.  (There exist schemes that perform
worse than this strategy, because of the feature of quantum systems
that every measurement that acquires information will uncontrollably
disturb the system.)  This scheme does \emph{not} lie within the set
of discriminate-and-reprepare schemes described above (it is not
described by an EBTP map) but we will nonetheless refer to it as
``classical.''

The average fidelity of this scheme is given by
\begin{equation}
   F_{\rm N} = 1-p\cos^2\theta \,.
   \label{eq:fn}
\end{equation}
This performance is plotted in Fig.~\ref{fig:fid_all}(b).  Clearly,
this scheme performs best for small amounts of noise ($p\simeq 0$)
and for input states with Bloch vectors that are near the $z$-axis
(which is invariant under the dephasing noise).  In some non-trivial
regions of the $(p,\theta)$ parameter space, in particular in the
range of low noise, this ``do nothing'' scheme outperforms the
optimal ``discriminate and reprepare'' scheme.

\section{Quantum Control}
\label{sec:quantum}

In the previous section, we presented control schemes based on
classical concepts.  However, using techniques that may lead to
optimal control schemes for a classical system may not necessarily
lead to optimal schemes for a quantum system.  As we will now
demonstrate, the above classical control strategies can be
outperformed by using a strategy based on quantum concepts.

We note that the two classical schemes presented in the previous
section lie at the extreme ends of a spectrum:  the
``discriminate-and-reprepare'' strategy achieved maximum information
gain and induced a maximum disturbance, whereas the ``do-nothing''
strategy achieved zero disturbance but produced zero information
gain. As demonstrated by Fuchs and Peres~\cite{Fuc96}, there exist
an entire range of generalized measurements that trade off
information gain and disturbance. A possible avenue for improvement
in our control schemes is to tailor the measurement in such a
way as to find a compromise, if one exists, between acquiring
information about the noise but not disturbing the system too much
as a result of the measurement.

In the following we re-express the noise process $\mathcal{E}_p$ in a
way that suggests a strategy for constructing such an improved
feedback protocol.

\subsection{Reexpressing the noise}

To develop an intuitive picture, we will make use of a
\emph{preferred ensemble} for the quantum operation $\mathcal{E}_p$
describing the noise.  That is, we use a decomposition of the
operation into different Kraus (error) operators than that given in
Eq.~(\ref{eq:erho}).  The resulting quantum operation
$\mathcal{E}_p$ describing the noise, however, is equivalent.

Consider the following quantum operation on a qubit, viewed on the
Bloch sphere: with probability $1/2$, the Bloch vector of the qubit
is rotated by an angle $+\alpha$ about the $z$-axis, and with
probability $1/2$ it is rotated by $-\alpha$ about the $z$-axis.
Rotations about the $z$-axis are described by the operator
\begin{equation}\label{eq:ZRotation}
    Z_{\alpha} = \mathrm{e}^{-i\alpha Z/2} = \cos(\alpha/2)I - i
    \sin(\alpha/2)Z \,,
\end{equation}
and the quantum operation is then
\begin{align}
    \mathcal{E}_\alpha(\rho) &= \tfrac{1}{2}Z_\alpha\rho
    Z_\alpha^\dag + \tfrac{1}{2}Z_{-\alpha}\rho
    Z_{-\alpha}^\dag \nonumber \\
    &= \sin^2(\alpha/2) (Z\rho Z) + \cos^2(\alpha/2) \rho \,.
    \label{eq:PreferredEnsemble}
\end{align}
Thus, this quantum operation is equivalent to the dephasing noise
$\mathcal{E}_p$, with $p = \sin^2(\alpha/2)$.

Viewing the noise operation $\mathcal{E}_p$ with this preferred
ensemble, it is possible to describe the noise as rotating the Bloch
vector of the state by $\pm\alpha$ with equal probability.  A
possible control strategy, then, would be to attempt to acquire
information about the direction of rotation ($\pm \alpha$) via an
appropriate measurement, and then to \emph{correct} the system based
on this estimate.  Loosely, we desire a measurement that determines
whether the noise rotated the state one way ($+\alpha$) or another
($-\alpha$). Then, based on the measurement result, we apply
feedback:  a unitary operation (rotation) that takes the state of
the system back to the desired axis.

A projective measurement, wherein the state of the system collapses
to an eigenstate of the measurement, does not meet these
requirements because such a measurement destroys the
distinguishability of the two possible states.  Instead, we consider
the use of a \emph{weak} measurement, with a measurement strength
chosen to balance the competing goals of acquiring information and
leaving the system undisturbed.  We now show that such a strategy is
possible, and that there is a non-trivial optimal measurement
strength for this task.

\subsection{Weak non-destructive measurements}

For our quantum control scheme, we will make use of a
type of measurement that satisfies two
key requirements: (1)~the strength of the measurement should be
controllable, i.e., we should be able to vary the trade-off between
information gain and disturbance (back-action); and (2)~the
measurement should be non-destructive,
which leaving the measured system in an appropriate quantum state
given by the desired collapse map.
Such weak non-destructive measurements have recently been developed and
demonstrated in single-photon quantum optical
systems~\cite{pryde:190402,ralph:012113}.

Using the preferred ensemble describing the noise,
Eq.~(\ref{eq:PreferredEnsemble}), we expect intuitively that this
weak measurement should be along the $y$-axis of the Bloch sphere in
order to provide information about which direction ($\pm\alpha$) the
system was rotated, \emph{without} acquiring information about which
initial state the system was prepared in. One suitable family of
POVMs consists of two operators given by $E_m = M_m^\dag M_m$, for
$m=0,1$, where $M_m$ are the measurement operators~\cite{Nie00}
\begin{align}\label{eq:MeasurementOperator1}
    M_0 &= \cos(\chi/2)|{+}i\rangle\langle{+}i| +
    \sin(\chi/2)|{-}i\rangle\langle{-}i| \,, \\
    \label{eq:MeasurementOperator2}
    M_1 &= \sin(\chi/2)|{+}i\rangle\langle{+}i| +
    \cos(\chi/2)|{-}i\rangle\langle{-}i| \,.
\end{align}
The strength of the measurement depends on the choice of the parameter
$\chi$. The eigenstates of $Y$ are $|{\pm} i\rangle \equiv (|0\rangle\pm
i|1\rangle)/\sqrt{2}$.
The probabilities of obtaining the measurement results $m=0,1$ for a
qubit in the state $\rho_{\rm in}$ are given by
\begin{equation}\label{eq:BornRule}
    p_m = \text{Tr}[E_m \rho_{\rm in}] \,,
\end{equation}
and the resulting state of the qubit immediately after the
measurement is
\begin{equation}\label{eq:PostMeasState}
    \rho_{\rm out}^{(m)} = \frac{M_m \rho_{\rm in}
    M_m^{\dag}}{p_m} \,.
\end{equation}

Consider the following two limits. If $\chi=\pi/2$ the two
measurement operators are the same and are proportional to the
identity. As a result the outcome probabilities are independent of
the state and the state of the signal is unaltered by the
measurement.  If $\chi=0$, a projective measurement on the signal is
induced: the signal state is projected onto the state
$|{{-}i}\rangle$ ($|{{+}i}\rangle$) when the measurement result is 0
(1). For $0 < \chi < \pi/2$, the resulting measurement on the signal
is non-projective but non-trivial.

\begin{figure}
  \begin{center}
   \includegraphics[width=.45\textwidth]{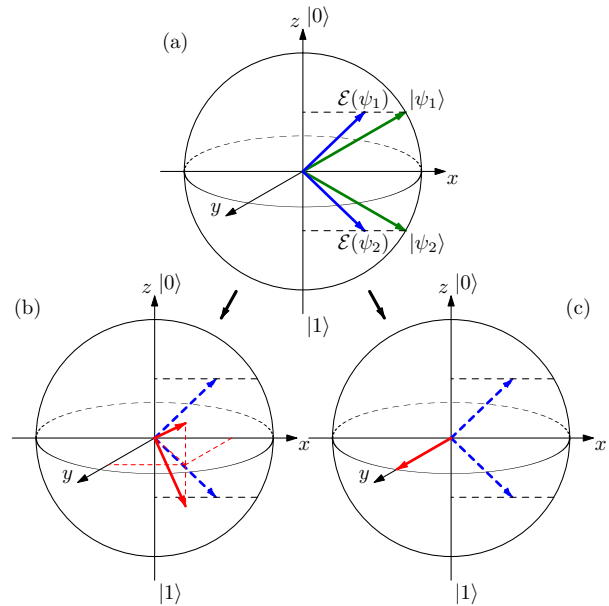}
  \end{center}
  \caption{Bloch sphere representation of the effect of a weak
  measurement on the system.  The transformations shown
  here correspond to having obtained the measurement result
  ``$0$''  (for the result ``$1$'',
  the behaviour would be a reflection in the $x$-$z$ plane.)
  a) The two initial states
  $|\psi_{1,2}\rangle$ are mapped to $\rho_{1,2}$ by the noise;
  b) a weak measurement is performed
  with $0<\chi<\frac{\pi}{2}$; c) a strong projective measurement
  ($\chi=0$) is performed projecting either state into
  $|{{-}i}\rangle$. While no measurement will not yield any information
  about the system,
  a strong measurement will maximally disturb the system.
  A weak measurement will gain some information
  while also limiting the disturbance on the system.}
  \label{fig:stateprog}
\end{figure}

It is illustrative to view the effect of this measurement on the
noisy input states on the Bloch sphere. In
Fig.~\ref{fig:stateprog}(a) we can see that the effect of the noise
is to shorten the length of the Bloch vector of the qubit state
(making it less pure) while \emph{increasing} the angle between the
Bloch vector and the $x$-$y$ plane from $\theta$ to $\theta'$, where
$\theta'>\theta$ . When the measurement is made, three things
happen, as can be seen in Fig.~\ref{fig:stateprog}(b): 1) the Bloch
vector is lengthened (the state becomes more pure); 2) the angle
$\theta'$ \emph{decreases} to some lesser angle $\theta''$; and 3)
the state is rotated about the $z$-axis one way or the other
depending on the result of the measurement.  The first two effects
work towards our advantage (purifying the state while decreasing
$\theta'$); the third effect we attempt to correct using
\emph{feedback}.

\begin{figure}
\begin{tabular}{l}
a)\\
\includegraphics[width=0.45\textwidth]{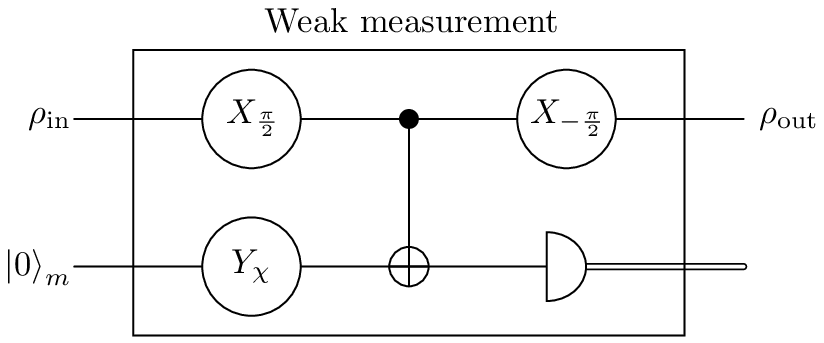}\\
b)\\
 \includegraphics[width=0.45\textwidth]{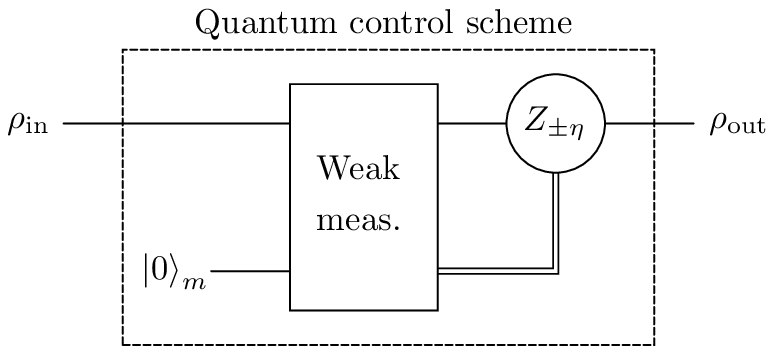}
\end{tabular}
\caption{a) Circuit diagram of the weak measurement scheme. The
input signal state $\rho_{\rm in}$ is entangled to the meter state
using the \textsc{cnot} gate. The $X_{\pm\frac{\pi}{2}}$ rotations
ensure that the weak measurement of the signal state is made in the
desired basis $\{|{+}i\rangle,|{-}i\rangle\}$. The strength of the
measurement is set using the rotation $Y_{\chi}$.  The meter state
is measured in the computational basis, resulting in a classical
signal (0 or 1) to be fed forward to the correction stage of the
control scheme. \\ b) Circuit diagram of the control scheme.  A weak
measurement is made on the input state and, based on the measurement
results, the signal state will be rotated by $Z_{\eta}$
($Z_{{-}\eta}$) conditional on the result of the weak measurement
being 0 (1).}
 \label{fig:schem}
\end{figure}

We will now describe how to implement this measurement using a
projective measurement on an ancillary {\it meter} qubit and an
entangling gate between the original {\it signal} qubit and the
meter. The strength of the measurement can be controlled by varying
the level of entanglement between the two qubits, which can be
implemented by initiating the meter in the state $|0\rangle$ and
subsequently applying a $Y_{\chi}$ rotation (as shown in figure
\ref{fig:schem}(a)), where
\begin{equation}
  Y_{\chi}=\mathrm{e}^{-i\chi
  Y/2}= \begin{pmatrix}\cos(\chi/2)&-\sin(\chi/2)\\
  \sin(\chi/2)&\cos(\chi/2)\end{pmatrix}\,.
\end{equation}
The parameter $\chi$ ranges from $0$ to $\pi/2$ and characterises
the strength of the measurement, with 0 equivalent to a projective
measurement and $\pi/2$ equivalent to no measurement.

The entangling gate consists of a $X_{\frac{\pi}{2}}$ rotation on
the signal state, followed by a \textsc{cnot} gate with the signal
state as the control and the meter state as the target, followed by
a $X_{-\frac{\pi}{2}}$ on the signal state, where
\begin{equation}
  X_{\phi}=\mathrm{e}^{-i\phi
  X/2}= \begin{pmatrix} \cos(\phi/2) & -i\sin(\phi/2)\\
  -i\sin(\phi/2)&\cos(\phi/2)\end{pmatrix} \,,
\end{equation}
and where the Pauli matrix $X$ is given by $X|0\rangle = |1\rangle$
and $X|1\rangle = |0\rangle$.  The rotations $X_{\pm\frac{\pi}{2}}$
are used to ensure that the resulting weak measurement on the signal
qubit is performed in the $\{|{+}i\rangle,|{-}i\rangle\}$ basis. The
entangling gate then correlates (to a degree which depends on
$\chi$) the $\{|{+}i\rangle,|{-}i\rangle\}$ basis of the signal
qubit to the $\{|0\rangle,|1\rangle\}$ basis of the meter qubit.

Finally the meter qubit is measured in the basis $\{|0\rangle,
|1\rangle\}$, yielding a result 0 or 1.  This measurement on the
meter induces a measurement on the signal that is precisely equal to
the generalized measurement described by the measurement operators
$M_m$ of Eq.~(\ref{eq:MeasurementOperator1}).

\subsection{Feedback control}

Once a weak measurement has been performed, a correction based on
the measurement result is performed on the quantum system: the
feedback control.  We choose the correction to be a unitary rotation
about the $z$-axis, $Z_{\pm\eta}$ where
\begin{equation}
  Z_{\eta}=\mathrm{e}^{-i\eta
  Z/2}= \begin{pmatrix} \mathrm{e}^{-i\eta/2}&0 \\
  0&\mathrm{e}^{+i\eta/2} \end{pmatrix} \,,
\end{equation}
with the aim to bring the Bloch vector of the qubit back onto the
$xz$-plane. The angle of rotation is chosen to be $\pm\eta$, depending on the
measurement result (${+}\eta$ corresponding to the measurement
result 0, and ${-}\eta$ to the measurement result 1). It is possible
to choose $\eta$ so that the system state is returned to the
$xz$-plane for all values of $p,\theta$ and $\chi$ and for both
measurement outcomes by choosing
\begin{equation}
  \label{eq:etaopt}
  \tan\eta=\frac{1}{(1-2p)\cos\theta\tan\chi}\,,
\end{equation}
with $\eta$ in the range $0 \leq \eta \leq \pi/2$.  This angle $\eta$ can be
calculated because the dephasing noise has been previously
characterised (i.e., $p$ is known).

The resulting weak measurement followed
by feedback is thus described by a quantum operation (a CPTP map)
$\mathcal{C}_{\rm QC}$ acting on a single qubit, given by
\begin{equation}\label{eq:QCasCPTP}
  \mathcal{C}_{\rm QC}(\rho) = (Z_{{+}\eta}M_0) \rho
  (Z_{{+}\eta}M_0)^\dag + (Z_{{-}\eta}M_1) \rho
  (Z_{{-}\eta}M_1)^\dag \,,
\end{equation}
where the measurement operators $M_m$ are given by
Eqs.~(\ref{eq:MeasurementOperator1}-\ref{eq:MeasurementOperator2}).

In summary, the quantum control scheme operates by performing a weak
measurement of the system and then correcting it based on the
results of the measurement, as in Fig.~\ref{fig:schem}b).  The weak
measurement is made by entangling an ancillary meter state with the
signal state using an entangling unitary operation, then performing
a projective measurement of the meter state.  The level of
entanglement depends on the input state of the meter, which is
controlled by a $Y_{\chi}$ rotation; this level of entanglement in
turn determines the strength of the measurement.  After measurement
of the meter, the signal state is altered due to the measurement
back-action.  To correct for this back-action, a rotation about the
$z$-axis is applied to the state, returning it back to the
$xz$-plane. To characterise how well the scheme works, we now
investigate the average fidelity.

\subsection{Performance}

The performance of this quantum control scheme, quantified by the
average fidelity~(\ref{eq:AverageFidelity}), is
\begin{equation}\label{eq:fid}
    F_{\mathrm{QC}} = \tfrac{1}{2}\left[ 1 +
    \sin^2\theta \sin \chi+\cos \theta \sqrt{1-
    (1-r_x^2)\sin^2\chi}\right]\,,
\end{equation}
where $r_x = (1-2p)\cos{\theta}$ is the $x$ component of the Bloch vector
describing the system after the noise.

\begin{figure}
  \begin{center}
   \includegraphics[width=.45\textwidth]{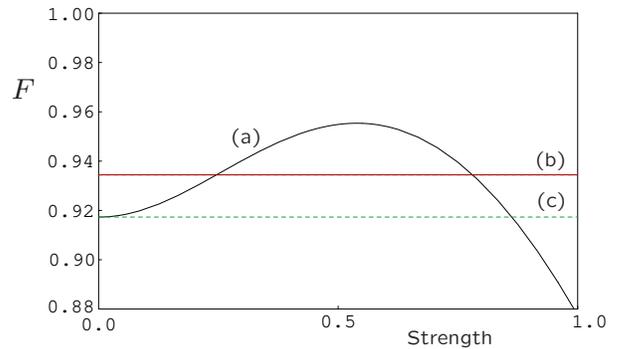}
  \end{center}
  \caption{(a) Fidelity of the quantum correction procedure with measurement
  strength ($1-2\chi/\pi$) for a representative noise value ($p=0.145$) and
  angle ($\theta=0.715$). The measurement strength ranges from a value of 0
  (corresponding to no measurement), through to a value of 1
  (corresponding to a projective measurement).
  There exists an optimum measurement strength at which we balance the amount
  of information gained with the amount of
  back-action noise introduced.  Also plotted for comparison
  are (b) the optimal ``discriminate-and-reprepare'' scheme and
  (c) the ``do nothing'' scheme for the same parameter values.}
  \label{fig:optimum_strength}
\end{figure}

We can see that $F_{\mathrm{QC}}$ is a function of the amount of
noise $p$, the angle between the initial states $\theta$, and the
measurement strength $\chi$.  The dependence of this fidelity on the
measurement strength, for fixed $p$ and $\theta$, is illustrated in
Fig.~\ref{fig:optimum_strength}.  For each value of $p$ and
$\theta$, there is an \emph{optimum} measurement strength
$\chi_{\mathrm{opt}}$ which maximizes the average
fidelity~(\ref{eq:fid}).  This optimum measurement strength is found
to be non-trivial except for the limiting cases of $p=0$ or
$\theta=0,\pi/2$, and is given by
\begin{equation}\label{eq:chiopt}
   \chi_{\rm opt}(p,\theta)
   \equiv \sin^{-1}\sqrt{\frac{\sin^4\theta}{(1-r_x^2)^2\cos^2\theta +
     (1-r_x^2)\sin^4\theta}}\,,
\end{equation}
as a function of the amount of noise $p$ and the angle between the
initial states $\theta$.

Substituting $\chi_{\mathrm{opt}}$ for $\chi$ in Eq.
(\ref{eq:fid}), we get the following expression for the optimum
fidelity:
\begin{equation}\label{eq:fid_qc}
    F_{\mathrm{QCopt}} = \tfrac{1}{2} +\tfrac{1}{2}
    \sqrt{\cos^2\theta +
    \frac{\sin^4\theta}{1-r_x^2}}\,.
\end{equation}

Fig.~\ref{fig:fid_all}(c) plots the quantum control fidelity as a
function of the input state (characterised by the angle $\theta$)
and the amount of noise (characterised by $p$).

We note that $F_{\mathrm{QCopt}} = 1$ for three limiting cases. If
$p=0$, there is no noise and so the state is not perturbed,
resulting in unit fidelity for all values of $\theta$ given by
simply ``doing nothing'' (zero measurement strength and no
feedback). When $\theta=\pi/2$, the states are orthogonal and point
along the $z$ axis.  The noise does not affect these states, again
resulting in unit fidelity for all values of $p$ with a ``do
nothing'' scheme. When $\theta=0$ the two states are equal and point
along the $x$-axis. The control scheme reprepares this state after
the noise by making a projective measurement $\chi=0$ to obtain
either $|{{+}i}\rangle$ or $|{{-}i}\rangle$ and rotating back to the
$xz$-plane ($\eta=\pi/2$). This results in a fidelity of $1$ for all
values of $p$.

\subsection{Comparison with Classical Schemes}

\begin{figure}
\begin{center}
\includegraphics[width=0.45\textwidth]{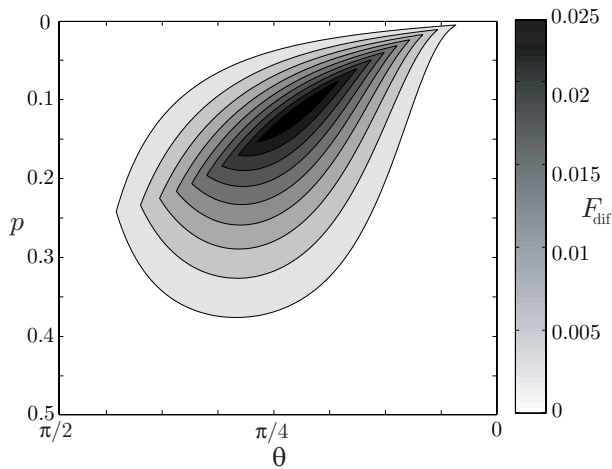}
\end{center}
\caption{A contour plot of the difference, as a function of the
amount of noise $p$ and the angle between the initial states
$\theta$, between the average fidelities of the quantum control
scheme and the best classical scheme.  The quantum control scheme
performs significantly better for moderate values of $p$ and $\theta$
($0.05\lesssim p\lesssim 0.3$ and
$0.3\lesssim \theta \lesssim 1$).
The maximum value $ F_{\mathrm{dif}}=0.026$ occurs at $p=0.115$ and
$\theta=0.715$.}
 \label{fig:fid_dif}
\end{figure}

We now compare the quantum control scheme with classical schemes
presented in Sec.~\ref{sec:Classical}.  Specifically, we compare the
quantum scheme with the best of the classical schemes at every point
in the parameter space $(p,\theta)$, i.e., we observe the difference
in the average fidelities
\begin{equation}
  F_{\mathrm{dif}}=F_{\mathrm{QCopt}}
  -\mathrm{max}(F_{\rm DR2},F_{\rm N})\,,
\end{equation}
where $F_{\rm DR2}$ and $F_{\rm N}$ are given by
Eqs.~(\ref{eq:fcl2}) and (\ref{eq:fn}), respectively.
Fig.~\ref{fig:fid_dif} reveals that $F_{\textrm{dif}}$ is always
positive, and thus the quantum control scheme always outperforms the
best of the classical strategies.

\subsection{Optimality}\label{sec:optimality}

We now prove that our quantum control scheme is optimal, in that it
yields the maximum average fidelity of all possible quantum
operations (CPTP maps).  Our proof makes use of techniques from
convex optimization (specifically, those of~\cite{audenaert2002})
but is presented without requiring any background in this subject.
In the Appendix, we provide a more detailed construction of the
proof.

Consider the following optimization problem: determine the maximum
average fidelity
\begin{equation}
  F_{\rm opt} = \max_{\mathcal{C}} F_\mathcal{C}
  = \max_{\mathcal{C}} \tfrac{1}{2}\sum_{i=1}^{2}{\langle\psi_i|
  \mathcal{C}\bigl[\mathcal{E}_p(|\psi_i\rangle\langle\psi_i|)\bigr]
  |\psi_i\rangle}\,,
  \label{eq:CPTPoptimisation}
\end{equation}
where the maximization is now over all CPTP maps $\mathcal{C}$
acting on a single qubit.

Recall that any CPTP map $\mathcal{C}$ acting on operators on a
Hilbert space $\mathcal{H}$ is in one-to-one correspondence with a
density operator $\Upsilon_{\mathcal{C}}$ on $\mathcal{H}\otimes
\mathcal{H}$ with
\begin{equation}
  \mathcal{C}(\varrho)={\rm
  Tr_{in}}\left[(\varrho^T\otimes I)\Upsilon_{\mathcal{C}}\right]\,,
\end{equation}
and is subject to the constraint ${\rm Tr}_{\rm out}
\left[\Upsilon_{\mathcal{C}}\right] = I_{\rm in}$, where `in' denotes the first
subsystem and `out' denotes the
second~\cite{72jamiolkowski275,Dar01,Nie00}. With this isomorphism,
the
average fidelity $F_\mathcal{C}$ for the control scheme $\mathcal{C}$ is given
by $F_\mathcal{C} = {\rm Tr}\left[R \Upsilon_{\mathcal{C}}\right]$, where
\begin{equation}
  R\equiv\tfrac{1}{2}\sum_{i=1}^2
  {\mathcal{E}_p\bigl(|\psi_i\rangle\langle\psi_i |\bigr)
  \otimes|\psi_i\langle\rangle \psi_i|}\,.
\end{equation}
Thus, the optimization problem (\ref{eq:CPTPoptimisation}) can be
rewritten as
\begin{equation}\label{eq:qprob_compact}
  \begin{array}{rl}
  \text{maximize}&{\rm Tr}\left[R \Upsilon_{\mathcal{C}}\right]\\
  \text{subject to}&\Upsilon_{\mathcal{C}}\geq 0\\
  &{\rm Tr}_{\rm out} \left[\Upsilon_{\mathcal{C}}\right] = I_{\rm in}
  \,.
  \end{array}
\end{equation}
We now wish to prove that the maximum value of ${\rm Tr}\left[R
\Upsilon_{\mathcal{C}}\right]$ subject to these constraints is given
by $F_{\rm QCopt}$ of Eq.~\eqref{eq:fid_qc}.

We note that, for any single-qubit operator $M$ satisfying $M\otimes
I - R \geq 0$, we obtain the inequality
\begin{align}
    {\rm Tr}\left[M\right] - {\rm Tr}\left[R \Upsilon_{\mathcal{C}}\right]
    &= {\rm Tr} \left[(M \otimes I)\Upsilon_{\mathcal{C}}\right] - {\rm Tr}
    \left[R \Upsilon_{\mathcal{C}}\right] \nonumber \\
    &= {\rm Tr} \left[(M \otimes I - R)\Upsilon_{\mathcal{C}}\right] \nonumber \\
    &\geq 0 \,,
\end{align}
where the first line follows from the constraint ${\rm Tr}_{\rm out}
\left[\Upsilon_{\mathcal{C}}\right] = I_{\rm in}$, and the
inequality follows from the fact that $(M \otimes I - R)\geq 0$ and
$\Upsilon_{\mathcal{C}} \geq 0$, and thus the trace of their product
is non-negative.  This inequality demonstrates that the value ${\rm
Tr}\left[M\right]$ for any matrix $M$ that satisfies the constraint
$(M \otimes I - R)\geq 0$ provides an upper bound on the solution of
our optimization problem~\eqref{eq:qprob_compact}.

Consider the matrix $M = b_0 (I + r_x X)$, where
\begin{equation}
  b_0=\frac{1}{4}
  +\frac{1}{4}\sqrt{\cos^2{\theta}+\frac{\sin^4{\theta}}{1-r_x^2}}\,,
\end{equation}
and $r_x = (1-2p)\cos\theta$ as before. It is straightforward to
verify that the matrix $b_0 I\otimes I+r_x b_0 X\otimes I - R \geq
0$, and hence the value ${\rm Tr}\left[M\right] = 2 b_0$ provides an
upper bound on the average fidelity of any control scheme. Because
$2 b_0$ precisely equals the fidelity of our proposed quantum
control scheme, given by Eq.~(\ref{eq:fid_qc}), this scheme
necessarily gives an optimal solution to the original problem
(\ref{eq:qprob_compact}). We refer the reader to the Appendix for a
more constructive proof of this result.

\section{Discussion and Conclusions}
\label{sec:conclusions}

We have shown how two key characteristics of quantum physics -- that
non-orthogonal states cannot be perfectly discriminated, and that
any information gain via measurement necessarily implies disturbance
to the system -- imply that classical strategies for control must be
modified or abandoned when dealing with quantum systems.  By making
use of more general measurements available in quantum mechanics, we
can design quantum control strategies that outperform schemes based
on classical concepts.  In particular, we have presented a task for
which the optimal scheme relies on a non-trivial measurement
strength, one that balances a tradeoff between information gain and
disturbance.

In constructing our quantum control scheme for the particular task
presented here, we made use of several intuitive guides.  First, we
used a preferred (and non-standard) ensemble of the dephasing noise
operator (Eq.~(\ref{eq:PreferredEnsemble})), which allowed us to
view the noise as ``kicking'' the state of the qubit in one
direction or the other on the Bloch sphere.  We then made use of a
weak measurement in a basis that, loosely, attempted to acquire
information about the direction of this kick without acquiring
information about the choice of preparation of the
system.  It is remarkable (and perhaps simply lucky) that these
intuitive guides lead to a quantum control scheme that was optimal
for the task.  It is interesting to consider whether such intuition
can be applied to quantum control schemes in general, and if this
intuition can be formalized into rules for developing optimal
control schemes.

While our scheme is indeed optimal for the task presented, it is not
guaranteed to be unique; in fact, there are other decompositions of
the same CPTP map into different measurements and feedback
procedures~\cite{BlumeKohoutCombes}. In general, it is possible that
an entire class of CPTP maps may yield the optimal performance.
Also, the intuitive guides discussed above for our quantum control
scheme --- such as that the measurement essentially gains
information only about the noise and not the choice of initial state
--- may not apply to other optimal schemes.

In connection to this, we note that a similar feedback control
scheme was investigated by Niu and Griffiths~\cite{Niu99} for
optimal eavesdropping in a B92 quantum cryptography
protocol~\cite{Ben92}, see also~\cite{Fuc96}.  In their scheme, the
aim of the weak
measurement was to \emph{maximize} the information gain about which
of two non-orthogonal states was transmitted for a given amount of
disturbance; in contrast, our weak measurement was designed to
acquire \emph{no} information about the choice of non-orthogonal
states. Despite these opposing aims, the obvious similarity between
these our scheme and that of Niu and Griffith warrants further
investigation, particularly since we note that optimal feedback
protocols exist based on different choices of measurement.

Finally, we note that the key element to our quantum control scheme
--- weak QND measurements on a qubit, and feedback onto a qubit based
on measurement results --- have both been demonstrated in recent
single-photon quantum optics experiments.  Specifically,
Pryde~\emph{et al}~\cite{pryde:190402} have demonstrated weak QND
measurements of a single photonic qubit, and have explicitly varied
the measurement strength over the full parameter range.  Also,
Pittman~\emph{et al}~\cite{pittman:052332} have demonstrated
feedback on the polarization of a single photon based on the
measurement of the polarization of another photon entangled with the
first; this feedback was used for the purposes of quantum error
correction, and is essentially identical to the feedback required
for our quantum control scheme.  Because these core essential
elements have already been demonstrated experimentally, we expect
that a demonstration of our quantum control scheme is possible in
the near future.

\begin{acknowledgments}
  We thank Sean Barrett, Robin Blume-Kohout, Jeremy O'Brien, Geoff Pryde,
  Kevin Resch, Andrew White, and
  Howard Wiseman for helpful discussions. P.E.M.F.M. acknowledges
  the support of the Brazilian agency Coordena\c c\~ao de
  Aperfei\c coamento de Pessoal de N\' ivel Superior (CAPES).
  This project was supported by the Australian Research Council.
\end{acknowledgments}

\appendix*

\section{Optimization Proofs}

In Sec.~\ref{sec:Classical} and \ref{sec:quantum}, the proposed
classical and quantum control schemes were shown to be optimal among
the set of EBTP and CPTP maps, respectively.  Here, we provide
constructive proofs of these results in further detail.

\subsection{Weak duality}

Consider the following optimization:
\begin{equation}\label{eq:dual}
  \begin{array}{rl}
  \text{maximize}&{\rm Tr}\left[F_0 Z\right]\\
  \text{subject to}&Z\geq 0\\
  &{\rm Tr}\left[F_i Z\right] = c_i
  \end{array}
\end{equation}
where the matrices $F_0$, $F_i$ and the vector $c$ are specific to
the problem, and $Z$ is the variable over which the optimization is
performed.  We say that any $Z$ satisfying the constraints of the
problem is {\it feasible}. An optimization problem of this form is
known as a \emph{semi-definite program} (SDP), a class of convex
optimization problems~\cite{Boy04}. Each problem of the form
(\ref{eq:dual}) has a Lagrange \emph{dual} optimization problem that
arises from using the method of Lagrange multipliers and has the
form~\cite{Boy04}
\begin{equation}
  \begin{array}{rl}\label{eq:ineq_form}
  \text{minimize}&c^T x\\
  \text{subject to}&-F_0 + \sum_{i}{x_i F_i} \geq 0
  \end{array}
\end{equation}
where now the vector $x$ is the variable to be optimized.

In many cases, such as the optimization problems investigated here,
the dual problem is straightforward to solve, or else efficient
numerical solutions are known that solve the primal and dual problems
together. Furthermore, the dual problem allows us to bound the
optimum of the original problem and this fact can be used to prove the
optimality of solutions as follows.

Let $\mathfrak{d}=c^T x$ be the value of the objective function to be minimized
in (\ref{eq:ineq_form}) for an arbitrary feasible $x$. Similarly, let
$\mathfrak{p}={\rm Tr}\left[F_0 Z\right]$ for an arbitrary feasible $Z$ and let
$\mathfrak{p}^*$ be the optimum of our original problem~(\ref{eq:dual}).

We now demonstrate that, if one can find a feasible point to
\eqref{eq:dual} yielding $\mathfrak{p}$ and a feasible point to
(\ref{eq:ineq_form}) yielding $\mathfrak{d}$ such that
$\mathfrak{d}=\mathfrak{p}$, then $\mathfrak{p}=\mathfrak{p}^*$,
that is, the point $Z$ yielding $\mathfrak{p}$ is optimal.

Consider the difference
\begin{equation}
  \mathfrak{d}-\mathfrak{p}=c^T x - {\rm Tr}\left[F_0 Z\right]
  ={\rm Tr}\Bigl[\Bigl(\sum_{i}{F_i x_i}-F_0\Bigl)Z\Bigr]\,,
\end{equation}
where we have used the linearity of the trace and $c^T x =
\sum_i{c_i x_i} =\sum_i{{\rm Tr}\left[F_i Z x_i\right]}$. As the
trace is over the product of two positive semi-definite matrices, it
has to be non-negative. That is to say that
\begin{equation}
  \mathfrak{p} \leq \mathfrak{p}^* \leq
  \mathfrak{d}\, .
\end{equation}
Clearly, if there is a $\mathfrak{p}$ such that
$\mathfrak{p}=\mathfrak{d}$ for some $\mathfrak{d}$, then
$\mathfrak{p}=\mathfrak{p}^*$.

\subsection{Dual optimization for quantum control}

As demonstrated in Sec.~\ref{sec:optimality}, obtaining the maximum
average fidelity can be expressed as the optimization problem
(\ref{eq:qprob_compact}).  For this problem (as for the classical
problem which we address in the next section) the dual optimization
proves to be straightforward to solve analytically and the results
above can then be used to show optimality of the control
scheme  given by Eq. (\ref{eq:QCasCPTP}).

We make use of some symmetry arguments to simplify the problem. This
optimization problem has certain symmetry properties under the
action of the group of transformations generated by the rotation
$\Upsilon\rightarrow (X\otimes X) \Upsilon (X\otimes X)^\dagger$ and
the transpose $\Upsilon\rightarrow \Upsilon^T$. Specifically, the
objective function is invariant under the action of this group since
${\rm Tr}[R (X \otimes X)\Upsilon (X \otimes X)]={\rm Tr}[R
\Upsilon]$ and ${\rm Tr}[R \Upsilon^T]={\rm Tr}[R \Upsilon]$,
because $(X \otimes X) R (X \otimes X) = R$ and $R^T=R$,
respectively.  In addition, the constraints are covariant under the
action of the group. Since conjugation with a unitary and
transposition preserve eigenvalues, $(X\otimes X) \Upsilon (X\otimes
X) \geq 0$ and $\Upsilon^T\geq 0$ if $\Upsilon \geq 0$. To see that
the equality constraints are covariant note that ${\rm Tr}_{\rm out}
\left[\Upsilon_{\mathcal{C}}\right] = I_{\rm
  in}$ is equivalent to the condition ${\rm Tr} \left[(M\otimes I_{\rm
    out}) \Upsilon \right]
= {\rm Tr} M$ for all hermitian $M$. If $\Upsilon$ obeys the partial
trace constraint we have
\begin{multline}
  {\rm Tr} [(M\otimes I_{\rm out})(X \otimes X) \Upsilon (X \otimes X)] \\
    = {\rm Tr} [(XMX\otimes I_{\rm out}) \Upsilon ] = {\rm Tr}M \,,
\end{multline}
and
\begin{equation}
    {\rm Tr} [(M\otimes I_{\rm out})\Upsilon^T ]
    = {\rm Tr} [(M^T\otimes I_{\rm out}) \Upsilon ]
    = {\rm Tr}M \,,
\end{equation}
so both $(X \otimes X) \Upsilon (X \otimes X) $ and $\Upsilon^T$ do also.
So both the objective
function and the feasible set of (\ref{eq:qprob_compact}) are
invariant under the action of the group. As a result there will be an
invariant point $\Upsilon^*_{\rm inv}=(X\otimes X)\Upsilon^*_{\rm inv}(X\otimes
X)=\Upsilon^{*T}_{\rm inv}$ that achieves the optimum
$\mathfrak{p}^*$~\cite{Boy04}. We do not need to optimize
over the full set of $\Upsilon$ but may restrict our attention to the
set of invariant
$\Upsilon_{\rm inv}$. Gatermann and Parrilo~\cite{Gat04} have
investigated such invariant SDP's in detail.

The dual of our optimization problem (\ref{eq:qprob_compact}) has the
form~\cite{audenaert2002}
\begin{equation}
  \begin{array}{rl}\label{eq:ineq_form2}
  \text{minimize}& \text{Tr} M \\
  \text{subject to}&M\otimes I - R \geq 0
  \end{array}
\end{equation}
Notice that (as is generally the case) this semidefinite program is
invariant under the same group of transformations as the original
problem, under which
$M\rightarrow XMX$ and
$M\rightarrow M^T$. For the dual problem we may likewise restrict
attention to $M_{\rm inv}= b_0I+b_x X$ that are invariant under the
action of the group. This gives a simpler dual optimization
\begin{equation}\label{eq:qdual}
  \begin{array}{rl}
  \text{minimize}&2 b_0\\
  \text{subject to}&b_0 I\otimes I+b_x X\otimes I - R \geq
  0 \,,
  \end{array}
\end{equation}
where $b_0$ and $b_x$ are the new variables.  This problem is simple
enough to solve analytically; the solution is
\begin{equation}
  b_0 =\frac{1}{4}+\frac{1}{4}\sqrt{\cos^2{\theta}+\frac{\sin^4{\theta}}{1-r_x^2}}\,,
\end{equation}
and $b_x = r_x b_0$ (with $r_x = (1-2p)\cos\theta$).  This may be
checked by verifying that the matrix $b_0 I\otimes I+b_x
X\otimes I - R$ is indeed positive semi-definite, hence $2 b_0$ is a
valid dual feasible value. Because $2 b_0$ reproduces the fidelity
of our proposed scheme, given by Eq. (\ref{eq:fid_qc}), this guess
necessarily gives an optimal solution to the original problem
(\ref{eq:CPTPoptimisation}).

\subsection{Dual optimization for classical control}

The same approach is used to solve the problem~(\ref{eq:EBTPoptimisation}). We
start by mapping the set of trace-preserving entanglement breaking qubit
channels to bipartite states $\Upsilon_{\mathcal{B}}$. For these channels
$\Upsilon_{\mathcal{B}}$ is positive, has partial trace equal to the identity,
and is also \emph{separable} \cite{HSR03}.  Because $\Upsilon_{\mathcal{B}}$ is an
(unnormalised) state of two qubits, the separability condition is equivalent to
the positivity of the partial transpose~\cite{horodecki1996a}. We will denote
the partial transpose of the operator $\Upsilon_{\mathcal{B}}$ on the subsystem
$\mathcal{H}_{\rm out}$ by $\Upsilon_{\mathcal{B}}^{T_{\rm out}}$ .
Thus we may rephrase the optimization
problem~(\ref{eq:EBTPoptimisation}) in the form
\begin{equation}\label{eq:prob_compact}
  \begin{array}{rl}
  \text{maximize}&{\rm Tr}\left[R\Upsilon_{\mathcal{B}}\right]\\
  \text{subject to}&\Upsilon_{\mathcal{B}} \geq 0\,,\quad \Upsilon_{\mathcal{B}}^{T_{\rm out}}\geq 0\\
  &{\rm Tr}_{\rm out} \Upsilon_{\mathcal{B}} = I_{\rm in}.
  \end{array}
\end{equation}
Note that the
condition of positivity of the partial transpose guarantees that
$\Upsilon_{\mathcal{B}}$ corresponds to an EBTP map.

The new problem has the same symmetries as the full optimization
(\ref{eq:qprob_compact}) with
one addition.   Notice that $R^{T_{out}}=R$ so the objective function
of both problems is invariant under partial transpose. In our new
problem the point $\Upsilon_{\mathcal{B}}^{T_{\rm out}}$ is feasible if $\Upsilon_{\mathcal{B}}$
is feasible, so the feasible set is also invariant under the partial
transpose. (Note that since partial transpose does not preserve
positivity this is not true of the problem
(\ref{eq:qprob_compact})). Because of this symmetry we may restrict
our attention to $\Upsilon_{\rm inv}$ for which $\Upsilon^{T_{\rm
    out}}_{\rm inv}=\Upsilon_{\rm inv}$. Since the partial transpose
sends $A\otimes Y \rightarrow -A\otimes Y$ where $A$ is any Hermitian
matrix, we can conclude that ${\rm
  Tr}[(A \otimes Y)\Upsilon_{\rm inv}]=0 $. It is sufficient to
check this condition for the full set of Pauli matrices
$I,X,Y,Z$ so the requirement of invariance under the partial transpose
constitutes four new constraints. Notice however that the condition
$\Upsilon^{T_{\rm out}}_{\rm inv}\geq 0$ is now redundant since we are
requiring that $\Upsilon^{T_{\rm out}}_{\rm inv}=\Upsilon^{T_{\rm
    inv}}$. So we can
replace the problem~(\ref{eq:prob_compact}) with
\begin{equation}\label{eq:prob_compact2}
  \begin{array}{rl}
  \text{maximize}&{\rm Tr}\left[R\Upsilon_{\mathcal{B}}\right]\\
  \text{subject to}&\Upsilon_{\mathcal{B}} \geq 0\\
  &{\rm Tr}_{\rm out} \Upsilon_{\mathcal{B}} = I_{\rm in} \\
& {\rm
  Tr}(A \otimes Y)\Upsilon_{\mathcal{B}}=0\quad \forall A \in \{I,X,Y,Z\}.
  \end{array}
\end{equation}
Positivity of the partial
transpose and hence the separability of $\Upsilon_{\mathcal{B}}$ is now guaranteed
by the positivity of $\Upsilon_{\mathcal{B}}$ and the additional
equality constraints.

The dual of the problem~(\ref{eq:prob_compact2}) is
\begin{equation}
  \begin{array}{rl}\label{eq:ineq_form_class}
  \text{minimize}& \text{Tr} M \\
  \text{subject to}&I\otimes M + N\otimes Y - R \geq 0
  \end{array}
\end{equation}
This semidefinite program still has symmetries corresponding to
the rotation $X\otimes X$ and the transpose (but not under the partial
transpose.) These two symmetries lead to the transformations
$N\rightarrow - XNX$ and $N\rightarrow -N^T$ respectively. The only
invariant choices of $N$ are proportional
to $Y$. As before we may restrict
attention to $M_{\rm inv}= a_0I+a_x X$ that are invariant under the
action of the group and $N_{\rm inv}=a_y Y$. This gives a simpler dual optimization
\begin{equation}\label{eq:cdual}
  \begin{array}{rl}
  \text{minimize}&2 a_0\\
  \text{subject to}&a_0 I\otimes I+a_x
  X\otimes I +a_y Y\otimes Y - R \geq
  0 \,,
  \end{array}
\end{equation}
where $a_0,a_x$ and $a_y$ are the new variables. This problem should
 be compared to the analogous dual optimization in the quantum case~(\ref{eq:qdual}).
 Again, this
problem can be solved analytically, yielding the solution
\begin{align}
  a_0 &=\frac{1}{4}+\frac{1}{4}\sqrt{\cos^2{\theta}+\sin^{4}{\theta}}\,,\\
  a_x &=\frac{r_x}{4}+\frac{r_x}{4}
  \frac{\cos^2{\theta}}{\sqrt{\cos^2{\theta}+\sin^{4}{\theta}}}\,,\\
  a_y &=-\frac{r_x}{4}
  \frac{\cos{\theta}\sin^2{\theta}}{\sqrt{\cos^2{\theta}+\sin^{4}{\theta}}}\,.
\end{align}
Again, one can check that $a_0 I\otimes I+a_x X\otimes I+a_y Y\otimes Y -
R$ is positive semidefinite with these choices, which ensures that
the objective function $2 a_0$ is indeed a dual feasible value. The proof
of optimality follows as before in the quantum case by: (i) observing
that $2 a_0$ reproduces the fidelity $F_{\rm DR2}$ of Eq.
(\ref{eq:fcl2}) and (ii) applying the weak duality argument.

We note that the optimization techniques presented here may be
useful when applied to more general problems presented in Fuchs and
Sasaki~\cite{Fuc03}.  However, when the map in question does not act
on qubits, there are significant complications in characterizing the
EBTP maps because the PPT condition is no longer sufficient.


\end{document}